\documentclass[aps,prx,groupedaddress,twocolumn,showpacs,longbibliography,10pt,nofootinbib]{revtex4-2}
\DeclareFontShape{OT1}{cmr}{mx}{n}{<->cmr10}{}
\fontseries{mx}\selectfont
\usepackage{color}
\usepackage{bm} 
\usepackage{units}
\usepackage{bbm}
\usepackage[caption=false]{subfig}
\usepackage{graphicx}
\usepackage{dsfont}
\usepackage{amsmath}
\usepackage{amssymb}
\usepackage{array}
\usepackage{epstopdf}
\usepackage{enumerate}
\usepackage{youngtab}
\usepackage{tensor}
\usepackage{braket}
\usepackage[normalem]{ulem}
\usepackage{slashed}
\usepackage[aligntableaux=center]{ytableau}
\usepackage[utf8]{inputenc}
\usepackage[
      colorlinks=true,
      linkcolor=blue,
     urlcolor=blue,
      filecolor=black,
      citecolor=red,
      ]{hyperref}
\usepackage{cleveref}
\usepackage{braket}
\usepackage{mathtools}
\usepackage{rotating}
\usepackage{tabularx}
\newcolumntype{Y}{>{\centering\arraybackslash}X}
\newcolumntype{C}[1]{>{\centering\arraybackslash}p{#1}}
\usepackage{todonotes}
\usepackage{xcolor}
\definecolor{LightCyan}{rgb}{0.7,1,1}
\definecolor{Gray}{gray}{0.9}
\usepackage{xspace}
\usepackage{slashed}

\begin{document}

\title{On the Entire Structure of the Energy Bands of 1D Moir\'{e} Superchain }

\author{Dmitrii Vorobev, Yiheng Chen and Grigory M. Tarnopolsky}
\affiliation{Department of Physics, Carnegie Mellon University, Pittsburgh, PA 15213, USA}

\begin{abstract}
We consider a general model of two atomic chains forming a moir\'{e} pattern due to a small mismatch in their lattice spacings, given by $\theta = (a_{1} - a_{2})/a_{2}$. Assuming arbitrary single-band dispersion relations $\varepsilon_{1}(p)$ and $\varepsilon_{2}(q)$ for the chains, along with an arbitrary inter-chain coupling term $T(x)$, we show that the entire spectrum of such a one-dimensional moir\'{e} superchain is governed by a single three-term recurrence  (TTR) relation. We  analyze this TTR relation using the discrete WKB method and demonstrate how the entire structure of the spectrum as well as emergence of flat bands can be easily identified from a pair of upper and lower potential functions of the TTR relation. We also comment on the chiral limit of the moir\'{e} superchain, which can be
viewed, in some sense, as a 1D analog of the chiral limit of Twisted Bilayer Graphene.

\end{abstract}

\maketitle
\nopagebreak

\section{Introduction}
The 2D moir\'{e} superlattices, as part of general van der Waals heterostructures, have recently become one of the mainstream research areas in modern condensed matter physics, both in experiments and theory \cite{CaoFatemiNature2018, CaoFatemiNature2, Yankowitz2018, Li2010, bistritzer2011moire, PhysRevB.82.121407, PhysRevX.8.031089, PhysRevB.98.075109, PhysRevB.98.085435, PhysRevB.98.035404, PhysRevB.98.045103, PhysRevLett.121.087001, PhysRevB.98.081102, PhysRevLett.121.257001, PhysRevB.99.075127,  PhysRevX.8.031088, Pizarro_2019, PhysRevX.8.031087, PhysRevB.98.241407, PhysRevX.8.041041, PhysRevB.98.235158, PhysRevB.98.220504,PhysRevB.99.144507, PhysRevB.106.235157, PhysRevLett.123.036401, PhysRevB.99.035111, PhysRevB.99.195455}.  Flat bands and magic angles in Twisted Bilayer Graphene (TBG)  are among the most fascinating examples of intriguing new physics emerging in these complex structures.
Recently, there has been progress in understanding the formation and properties of flat bands in TBG \cite{PhysRevLett.108.216802, PhysRevLett.122.106405, PhysRevResearch.2.023237, PhysRevB.103.155150, PhysRevResearch.3.023155, PhysRevLett.127.246403, PhysRevX.13.021012} and Twisted Graphene Multilayers \cite{PhysRevB.100.085109, PhysRevLett.123.026402, CeaWaletGuinea2019, PhysRevLett.125.116404, PhysRevLett.123.026402, lin2022energetic, ma2023doubled, PhysRevLett.123.026402, PhysRevB.105.195422, PhysRevB.100.085109, PhysRevLett.128.176404, PhysRevB.107.125423, PhysRevB.108.L081124, PhysRevResearch.6.L022025, Devakul2023, PhysRevX.13.041007, PhysRevResearch.5.043079, PhysRevB.109.205411}. Nevertheless, a general principle for analyzing and predicting the formation of flat bands in moiré lattices is not yet well understood.

In this article, we take a step toward a better understanding structure of the moir\'{e} bands by studying a simple yet general model of two one-dimensional chains of atoms with a small mismatch $\theta = (a_{1} - a_{2})/a_{2}$ between their lattice spacings $a_{1}$ and $a_{2}$, which together form a moir\'{e} superchain. This type of 1D superchain has been studied in many previous works \cite{Canc_s_2017, PhysRevResearch.2.033162, PhysRevB.101.041112, PhysRevB.101.041113, Tritsaris2021, Gon_alves_2022, PhysRevResearch.4.043224, PhysRevB.106.075420, PhysRevResearch.4.L032031, Dams2023, Schleder_2023, PhysRevB.107.224206} and a special case of the model we investigate was carefully analyzed in \cite{PhysRevResearch.2.033162},  from the perspective of the local density of states.
In this article, we show that the entire structure of the moiré  bands of such a superchain is governed by a single three-term recurrence (TTR) relation. The small mismatch parameter $\theta \ll 1$ enables the application of the discrete Wentzel–Kramers–Brillouin (WKB) method \cite{Braun1993}.  This method reveals that the band structure can be effectively analyzed using a pair of upper and lower potential functions of the TTR relation. 
We also note that a closely related 2D problem involves a 1D moir\'{e} pattern formed in the 2D superlattices due to lattice strain \cite{PhysRevLett.108.216802, PhysRevLett.125.166803, Becker2024, PhysRevB.107.235143}.

The article is organized as follows. In Section 2, we describe our model and derive its continuum Hamiltonian. In Section 3, we show how the moir\'{e} bands can be obtained by solving a single TTR relation. In Section 4, we demonstrate how the discrete WKB method reveals the entire structure of the moir\'{e} bands. Finally, in Section 5, we derive the semiclassical energy quantization conditions and explain the emergence of flat bands in certain energy regions of the spectrum.

\section{Continuum model for moir\'{e} superchain}
\label{Sec2}
\begin{figure}[h!]
\includegraphics[width=0.25\textwidth]{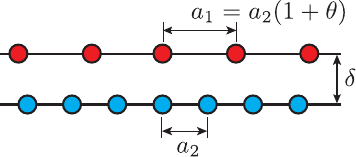}
\caption{Two one-dimensional chains of atoms with slightly different lattice spacings $a_{1}$ and $a_{2}$ form a moir\'{e} superchain. An electron can hop along each chain and between them. }
\label{model_pic1}
\end{figure}

Let us consider an electron that can hop along two 1D chains of atoms as well as between them, as shown in the Figure \ref{model_pic1}.
We assume that individual chains  have Hamiltonians $H_{1}$ and $H_{2}$ with  single-band dispersion relations $\varepsilon_{1}(p)$ and $\varepsilon_{2}(q)$, respectively. The inter-chain coupling term is given by $T(x_{n}-y_{m})$, where $x_{n}= n a_{1}$ denotes the coordinate of atom $n$ in the first chain, and $y_{m}=m a_{2}$ denotes the coordinate of atom  $m$ in the second chain. The Hamiltonian for such an electron is given by
\begin{align}
H =  H_{1}+H_{2} + \sum_{n,m}(T(x_{n}-y_{m}) c_{n}^{\dag}d_{m}+\textrm{H.c.})\,, \label{Ham0}
\end{align} 
where $c_{n}^{\dag}, c_{n}$ and $d^{\dag}_{m}, d_{m}$ are the electron creation and annihilation operators,  $a_{1}=a_{2}(1+\theta)$ and $\theta$ encodes a lattice mismatch between the chains and is the 1D equivalent to the twist angle in TBG. As an example we can consider the interchain hopping term $T(x)$ in one of the following forms:
\begin{align}
 T(x) = w e^{-|x|/\xi}, \quad 
T(x) = w e^{(\delta-\sqrt{x^{2}+\delta^{2}})/\xi}\,, \label{Txcases}
\end{align} 
or $T(x) =w e^{-x^{2}/\xi^{2}}$ as in \cite{PhysRevResearch.2.033162, Gon_alves_2022},  where $\xi$ is the correlation length  and $\delta$ is a distance between the chains.
We introduce creation and annihilation operators $c^{\dag}_{p}, c_{p}$ and $d_{q}^{\dag}, d_{q}$ in the momentum space using the formulas 
\begin{align}
c_{p} = \frac{1}{\sqrt{b_{1}}}\sum_{n} c_{n} e^{-i p x_{n} }, \, d_{q} =\frac{1}{\sqrt{b_{2}}} \sum_{m} d_{m} e^{-i q y_{m} }\,,
\end{align} 
where $b_{1}=2\pi/a_{1}$ and $b_{2}=2\pi/a_{2}$ are the reciprocal lattice primitive vectors. Using the momentum space operators we find for the Hamiltonian 
\begin{align}
H =&  \int_{0}^{b_{1}}dp \, \varepsilon_{1}(p) c^{\dag}_{p}c_{p} +\int_{0}^{b_{2}} dq \varepsilon_{2}(q)\, d^{\dag}_{q}d_{q} \notag\\
&+
\int_{0}^{b_{1}} dp \int_{0}^{b_{2}} dq (T_{p,q}c^{\dag}_{p}d_{q} + \textrm{H.c.})\,,
\end{align} 
and we  denoted by $T_{p,q}$ the expression 
\begin{align}
T_{p,q}= \frac{1}{\sqrt{b_{1}b_{2}}}\sum_{n,m}T(x_{n}-y_{m})e^{-ipx_{n}+iqy_{m}}\,.
\end{align} 
Let us define the continuous Fourier transform for the interchain hopping function $T(x)$ as 
\begin{align}
T(p) \equiv \frac{1}{\sqrt{a_{1}a_{2}}}\int_{-\infty}^{+\infty} dx\, T(x) e^{-ipx}\,,\label{FT0}
\end{align} 
and using this definition we obtain an expression for $T_{p,q}$
\begin{align}
T_{p,q}&=\sum_{G_{1},G_{2}} T(p-G_{1})\delta(q-p + G_{1}+G_{2})\,,
\end{align} 
where $G_{1} = m_{1}b_{1} $ and $G_{2} = m_{2}b_{2}$ are reciprocal lattice vectors of the first and second chains. Therefore the Hamiltonian  reads 
 \begin{align}
H &=  \int_{0}^{b_{1}}dp\, \varepsilon_{1}(p) c^{\dag}_{p}c_{p} +\int_{0}^{b_{2}}dq \, \varepsilon_{2}(q) d^{\dag}_{q}d_{q} \notag\\
&\; +
\int_{0}^{b_{1}}dp\, (\sum'_{G_{1},G_{2}} T(p-G_{1})c^{\dag}_{p}d_{p-G_{1}-G_{2}} + \textrm{H.c.})\,, \label{Hfinal0}
\end{align} 
where $\sum'_{G_{1},G_{2}}$ indicates that we sum only over those $G_{1}$ and $G_{2}$ for which $p - G_{1} - G_{2}$ lies within the interval $\textrm{BZ}_{2} = [0, b_{2}]$, the Brillouin zone for the $q$ momentum.

\begin{figure}[h!]
\includegraphics[width=0.35\textwidth]{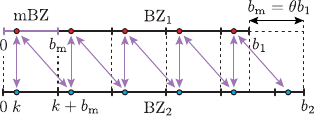}
\caption{Two Brillouin zones $\textrm{BZ}_{1}$ and $\textrm{BZ}_{2}$ of the first and second chains and the moir\'{e}  vector $b_{\textrm{m}} \equiv b_{2}-b_{1}$.  The moir\'{e} Bloch vector $k$ belongs to the moir\'{e} Brillouin zone $\textrm{mBZ}=[0,b_{\textrm{m}}]$ and the momenta $k+jb_{\textrm{m}}$, $j=0,1,\dots, n_{\textrm{m}}$ of the $\textrm{BZ}_{1}$ and $\textrm{BZ}_{2}$ are coupled by the continuum Hamiltonian (\ref{ContHam}).  }
\label{twoBZs0}
\end{figure}

Without loss of generality we assume that $a_{2}<a_{1}$, and thus $b_{2}>b_{1}$. We also assume that the function $T(p)$ decays rapidly with $p$ and has its maximum at $p = 0$. Therefore, $T(p - G_1)$ takes its largest values only for $G_1 = 0$ and $G_1 = -b_1$, so these are the terms we retain in the continuum model Hamiltonian
 \begin{align}
H_{\textrm{cont}} =& \int_{0}^{b_{1}}dp\, \varepsilon_{1}(p) c^{\dag}_{p}c_{p} +\int_{0}^{b_{2}}dq\, \varepsilon_{2}(q) d^{\dag}_{q}d_{q} \notag\\
&+\sum_{G_{2}}'\int_{0}^{b_{1}}dp\,\big(
  T(p)c^{\dag}_{p}d_{p-G_{2}} \notag \\
 &\quad+ T(p-b_{1})c^{\dag}_{p}d_{p-b_{1}-G_{2}}+ \textrm{H.c.}\big)\,.\label{Hcont0}
\end{align} 
Finally, since $p - G_{2}$ in the first interchain term must belong to the interval $\textrm{BZ}_{2} = [0, b_{2}]$, and since $p \in \textrm{BZ}_{1} = [0, b_{1}]$, we must take only $G_{2} = 0$. Similarly, $p - b_{1} - G_{2}$ in the second term lies in the interval $\textrm{BZ}_{2} = [0, b_{2}]$ only if $G_{2} = -b_{2}$. Thus, we finally find
\begin{align}
&H_{\textrm{cont}} =\int_{0}^{b_{1}}dp\, \varepsilon_{1}(p) c^{\dag}_{p}c_{p} +\int_{0}^{b_{2}}dq\, \varepsilon_{2}(q) d^{\dag}_{q}d_{q}  \notag\\
&+\int_{0}^{b_{1}}dp\,\big(
 T(p)c^{\dag}_{p}d_{p} + T(p-b_{1})c^{\dag}_{p}d_{p+b_{\textrm{m}}} + 
 \textrm{H.c.}\big)\,, \label{ContHam}
\end{align} 
where we denoted  by $b_{\textrm{m}} \equiv b_{2}-b_{1}$ the moir\'{e}  vector, depicted in Figure \ref{twoBZs0}.
We notice that $b_{2}=b_{1}(1+\theta)$ and therefore $b_{\textrm{m}}=\theta b_{1}$. The continuum model (\ref{ContHam}) is accurate, provided that the interchain hopping term $T(p)$ decays rapidly on the scale of $b_1$, so that the terms neglected in the sum over $G_1$ in (\ref{Hfinal0}) are small.

We see that the momenta $k+jb_{\textrm{m}}$, with $j=0,1,2,\dots$ of the $\textrm{BZ}_{1}$ and  $\textrm{BZ}_{2}$ are coupled to each other by the continuum model Hamiltonian (\ref{ContHam}) as shown in the Figure \ref{twoBZs0}. In general $b_{1} = n_{\textrm{m}}b_{\textrm{m}}+rb_{\textrm{m}}$, where $n_{\textrm{m}}$ is a  number of full moir\'{e} Brillouin zones $\textrm{mBZ}=[0,b_{\textrm{m}}]$ inside $\textrm{BZ}_{1}$ and  $0\leq r < 1$ is the reminder. If $r>0$, then for the first interval  $k\in [0, r b_{\textrm{m}})$ of $\textrm{mBZ}$ we have $j = 0,1,2,\dots,  n_{\textrm{m}}$ and for the second interval 
$k\in [r b_{\textrm{m}}, b_{\textrm{m}})$ we have $j = 0,1,2,\dots,  n_{\textrm{m}}-1$ for $\textrm{BZ}_{1}$ and 
$\textrm{BZ}_{2}$ has always one extra mode.  For simplicity below we assume a commensurate case, when  there are exactly integer number $n_{\textrm{m}}$ of the moir\'{e} vectors $b_{\textrm{m}}$ inside $b_{1}$, so the reminder $r=0$. In this case the moir\'{e} Hamiltonian is  a block matrix of the form
\begin{align}
&H_{\textrm{moir\'{e}}}(k) =  \left(  \begin{array}{cc}
    H_{1}(k) & T(k) \\ 
    T^{\dag}(k) & H_{2}(k) \\ 
  \end{array}\right)\,, \label{Hmoire}
\end{align} 
where $k\in \textrm{mBZ}=[0,b_{\textrm{m}}]$ and $H_{1}(k)_{j,j'} = \varepsilon_{1}(k + jb_{\textrm{m}}) \delta_{j,j'}$ with $j=0,\dots, n_{\textrm{m}}-1$ and  $H_{2}(k)_{j,j'} = \varepsilon_{2}(k + jb_{\textrm{m}}) \delta_{j,j'}$ with $j=0,\dots, n_{\textrm{m}}$ are diagonal matrices representing intrachain dispersion relations and $T(k)$ is  $n_{\textrm{m}}\times (n_{\textrm{m}}+1)$ interchain hopping matrix that has only diagonal and upper diagonal non-zero elements: 
$T(k)_{j,j} = T(k + jb_{\textrm{m}})$ and $T(k)_{j,j+1} = T(k + jb_{\textrm{m}}-b_{1})$. 

We notice that if we consider the limit when the intrachain terms are zero, $H_{1} = H_{2} = 0$, then the Hamiltonian (\ref{Hmoire}) has chiral symmetry and hosts an exact flat band at zero energy.  
This flat band arises from the fact that the matrix 
$T$ has one more column than rows, and thus has exactly one zero mode for each $k$. This limit can be viewed, in some sense, as a 1D analog of the chiral limit of TBG \cite{PhysRevLett.108.216802, PhysRevLett.122.106405}. We demonstrate below that this flat band is stable under small nonzero intrachain terms $H_{1}$ and $H_{2}$, while the discrete WKB method does not capture it.

\section{Moir\'{e} bands and three-term recurrence relation}
\label{Sec3}

To find the moir\'{e}  bands we solve the static Schr\"{o}dinger equation 
\begin{align}
&H_{\textrm{moir\'{e}}}(k) |\psi_{n}(k)\rangle = \varepsilon_{nk} |\psi_{n}(k)\rangle\,, \label{moireEigEq}
\end{align} 
for every $k\in \textrm{mBZ}=[0,b_{\textrm{m}}]$, where $n$ is the band index. 
The equation (\ref{moireEigEq}) can be written explicitly in components $\textbf{u} =(u_{0},\dots,u_{n_{\textrm{m}-1}})$ and $\textbf{v} =(v_{0},\dots,v_{n_{\textrm{m}}})$ of the 
vector $|\psi(k)\rangle = (\textbf{u}, \textbf{v})^{\textrm{T}}$, as
\begin{align}
\begin{cases}
\varepsilon_{1,j} u_{j} + t_{j} v_{j} + \tilde{t}_{j}v_{j+1}= \varepsilon u_{j}\\
\varepsilon_{2,j} v_{j} + t_{j}u_{j}  + \tilde{t}_{j-1}u_{j-1}= \varepsilon v_{j}\\
\end{cases}\,, \label{MoireSysEq}
\end{align} 
where we suppressed indices $n$ and $k$ for brevity and also introduced short notations $\varepsilon_{1,2; j} \equiv  \varepsilon_{1,2 }(k + jb_{\textrm{m}})$ and 
\begin{align}
 t_{j}\equiv T(k + jb_{\textrm{m}}), \quad  \tilde{t}_{j}\equiv T(k + jb_{\textrm{m}}-b_{1})\,.
\end{align} 
Using the first equation in (\ref{MoireSysEq}) to express $u_{j}$ in terms of $v_{j}$ and $v_{j+1}$, and 
substituting it into the second equation, we obtain a single equation for $v_{j}$ in the form
\begin{align}
\mu_{j}v_{j-1} + w_{j}v_{j} + \mu_{j+1}v_{j+1} =0\,, \label{TTR0}
\end{align} 
where $j=0,\dots, n_{\textrm{m}}$; we assume $v_{-1}=v_{n_{\textrm{m}}+1}=0$ and we introduced $\mu_{j} \equiv t_{j-1}\tilde{t}_{j-1}/(\varepsilon - \varepsilon_{1,j-1}) $ and 
\begin{align}
w_{j} \equiv \frac{t_{j}^2}{\varepsilon - \varepsilon_{1,j}} +\frac{\tilde{t}_{j-1}^2}{\varepsilon -\varepsilon_{1,j-1}} - (\varepsilon-\varepsilon_{2,j})\,.
\end{align} 
Equivalently one can obtain a similar equation for the components $u_{j}$.
The equation (\ref{TTR0}) is called a three-term recurrence (TTR) relation and is ubiquitous in physics and mathematics.  This TTR  can also be represented as an eigenvalue equation with a zero eigenvalue for a finite Hermitian Jacobi (tridiagonal) matrix \cite{Gantmacher1961OscillationMA}. 

For the moir\'{e} superchain we assume that the lattice mismatch parameter $\theta =b_{\textrm{m}}/b_{1}\ll 1$ is small. In this case, the coefficients $\mu_{j}$ and $w_{j}$ vary slowly with $j$, which allows us to apply the discrete WKB method \cite{Harper_1955, SchultenGordon1975, Sazonov1978, Braun1978, Braun1979, PhysRevLett.43.1954, Wilkinson1984, Braun1993} to the TTR relation (\ref{TTR0}) and to determine the entire moir\'{e} spectrum $\varepsilon$ both qualitatively and quantitatively.

\section{Entire structure of the moir\'{e} bands}
\label{Sec4}
In this section, we rely mostly on the  results and some notations presented in \cite{Braun1978, Braun1993}.  We first discuss qualitatively the structure 
of the energy levels determined by the TTR relation  (\ref{TTR0}).
At the leading level of accuracy we ignore $O(\theta)$ order corrections, and therefore can make approximations $\varepsilon_{1,j-1}\approx \varepsilon_{1,j} $ and $\tilde{t}_{j-1}\approx \tilde{t}_{j}$ in the coefficients $\mu_{j}$ and $w_{j}$ and we also take the Bloch vector $k=0$. This allows to rewrite  (\ref{TTR0}) in a simpler form
\begin{align}
\tilde{\mu}_{j}v_{j-1} + \tilde{w}_{j}v_{j} + \tilde{\mu}_{j+1}v_{j+1} =0\,, \label{TTR1}
\end{align} 
where $\tilde{\mu}_{j}\equiv t_{j}\tilde{t}_{j}$ and $\tilde{w}_{j}=t^{2}_{j}+\tilde{t}_{j}^{2} - (\varepsilon -\varepsilon_{1,j})
 (\varepsilon -\varepsilon_{2,j})$. Solving the equations $\tilde{w}_{j}\pm 2 \tilde{\mu}_{j}=0$ with respect to energy $\varepsilon$ we obtain two lower $U^{-}_{1,j}$, $U^{-}_{2,j}$ and two upper $U^{+}_{1,j}$, $U^{+}_{2,j}$ potential functions for the energy $\varepsilon$:
\begin{align}
U^{\pm }_{1,j}= \frac{\varepsilon_{1,j}+\varepsilon_{2,j}}{2} - \sqrt{\frac{(\varepsilon_{1,j}-\varepsilon_{2,j})^2}{4}+(t_{j}\mp \tilde{t}_{j})^2}\,, \\
U^{\pm }_{2,j}= \frac{\varepsilon_{1,j}+\varepsilon_{2,j}}{2} + \sqrt{\frac{(\varepsilon_{1,j}-\varepsilon_{2,j})^2}{4}+(t_{j}\pm\tilde{t}_{j})^2}\,,
\end{align} 
and we notice that for any $j$ the potential functions obey the chain of inequalities
\begin{align}
U_{1,j}^{-}<U_{1,j}^{+} \leq \varepsilon_{1,j}, \varepsilon_{2,j} \leq U_{2,j}^{-}<U_{2,j}^{+}\,. \label{ChainOfIneq}
\end{align} 
For a given energy $\varepsilon$, the classically allowed regions of $j$ are  determined by the inequalities 
\begin{align}
U^{-}_{1,j} < \varepsilon < U^{+}_{1,j},\quad U^{-}_{2,j} < \varepsilon < U^{+}_{2,j} \,,
\end{align} 
where, from now on, we treat the discrete variable $j$ as a continuous one. 
Here, we can already notice the difference from the usual WKB approach, in which a single potential function $U(x)$ defines the classically allowed and forbidden regions of a particle's motion.  In the discrete WKB method, the momentum operator $\hat{p} = -i \partial/\partial j$ enters the TTR relation as $2\mu_{j+1/2} \cos \hat{p}$ at the $O(\theta)$ order of accuracy, which leads to the upper and lower potential functions \cite{Braun1993}. 
We note that, in the usual TTR relation, the energy $\varepsilon$ enters the equation linearly \cite{Braun1993}. 
However, in our case, $\varepsilon$ enters the relation nonlinearly through the coefficients $\mu_{j}$ and $w_{j}$, 
giving rise to a pair of lower and upper potential functions. 
Nevertheless, the discrete WKB method remains conceptually unchanged.

The intervals of possible values of quantized energies $\varepsilon$ are bound by minimums of $U^{-}_{1,j}$ and $U^{-}_{2,j}$ and maximums of $U^{+}_{1,j}$ and $U^{+}_{2,j}$:
\begin{align}
&\textrm{min}_{j} \,U_{1,j}^{-} \leq \varepsilon \leq \textrm{max}_{j} \,U_{1,j}^{+}\,, \notag\\
&\textrm{min}_{j} \,U_{2,j}^{-} \leq \varepsilon \leq \textrm{max}_{j} \,U_{2,j}^{+} \,.\label{BoundsEnr}
\end{align} 
In fact one can rigorously prove \cite{Braun1979} that quantized energies $\varepsilon$ must satisfy the inequalities
\begin{align}
&\textrm{min}_{j}(w_{j}-|\mu_{j}|-|\mu_{j+1}|) \leq 0 \notag \\
&\textrm{max}_{j}( w_{j}+|\mu_{j}|+|\mu_{j+1}|) \geq 0 \,,
\end{align}
which agree with (\ref{BoundsEnr}) at the order $O(\theta)$. 

\begin{figure}[h!]
\includegraphics[width=0.48\textwidth]{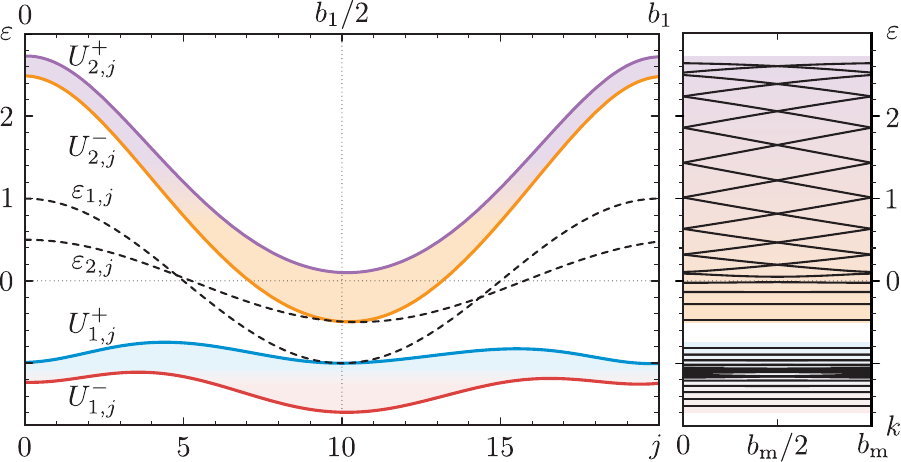}
\caption{(Left) Plot of the potential functions $U^{-}_{1,j},U^{+}_{1,j}$ and $U^{-}_{2,j},U^{+}_{2,j} $ (solid colored lines) and dispersions $\varepsilon_{1,j}$ and $\varepsilon_{2,j}$ (black dashed lines) for the case $\varepsilon_{1}(p) =\cos(p a_{1})$, $\varepsilon_{2}(q) =0.5\cos(q a_{2})$ and $T(p)$ in (\ref{Tpcase1}) with $w=3/2$, $\xi/a_{1}=0.6$ and $\theta =0.05$, with highlighted classically allowed regions of energy. (Right) Moir\'{e} bands obtained from exact numerical solution of (\ref{TTR0}).  }
\label{UpmAndBands1}
\end{figure}

As an example in Figure \ref{UpmAndBands1} 
we plot $U^{-}_{1,j}, U^{+}_{1,j}$ and $U^{-}_{2,j}, U^{+}_{2,j}$ functions together with the moir\'{e} bands obtained from exact numerical solution of  (\ref{TTR0}),   for a particular case when $\varepsilon_{1}(p) =\cos(p a_{1})$, $\varepsilon_{2}(q) =0.5\cos(q a_{2})$ and 
\begin{align}
T(p)=2w\sqrt{1+\theta}\frac{\xi/a_{1}}{1+p^2 \xi^2}\,,  \label{Tpcase1}
\end{align}
which is the Fourier transform (\ref{FT0}) of the interchain hopping function $T(x)=w e^{-|x|/\xi}$  with $w=3/2$ and $\xi/a_{1}=0.6$ and $\theta =0.05$. We notice that the mismatch parameter $\theta$ enters inter-chain hopping function $T(p)$ via the factor $1/\sqrt{a_{1}a_{2}}$ in (\ref{FT0}). We also highlighted the classically allowed regions of energy. We note that they coincide with the plots of the local density of states in momentum space for $\theta =0$, presented in \cite{PhysRevResearch.2.033162}.
 One can clearly see from  the Figure \ref{UpmAndBands1} that the exact moir\'{e} spectrum obeys inequalities (\ref{BoundsEnr}). We can also observe several regions of highly flat bands as well as the regions where bands strongly depend on the moir\'{e} Bloch vector $k$. We explain it in the next section by analysing solution of the TTR relation (\ref{TTR0}) using the Bohr–Sommerfeld quantization conditions for the discrete WKB method.

\section{Moir\'{e} bands from the discrete WKB method}
\label{Sec5}

To apply the discrete WKB method to the TTR relation (\ref{TTR0}) we treat the discrete variable $j$ as continuous one. Following \cite{Braun1978, Braun1993} we define two types of the turning points: a "usual" turning point $j_{t}$ is defined as the point where $\varepsilon = U^{+}_{j_{t}} $ and an "unusual" turning point corresponds to $\varepsilon = U^{-}_{j_{t}}$. In our case since the energy $\varepsilon$ enters TTR implicitly through $\mu_{j}$ and $w_{j}$ we find the left and right turning points $j_{\textrm{l}}$ and $j_{\textrm{r}}$ by solving the equations:
\begin{align}
w_{j} \pm  2 |\mu_{j+1/2}| = 0
\end{align}
with respect to $j$ for a fixed energy $\varepsilon$.\footnote{We emphasize again that we treat $j$ as a continuous variable and that $j_{\textrm{l}}$ and  $j_{\textrm{r}}$ are real numbers (not necessarily integers).} Now we are working at the accuracy $O(\theta)$, therefore we can not make approximations similar to those when we obtained (\ref{TTR1}).
The  Bohr–Sommerfeld quantization rules for the discrete WKB method depend on the type of the turning points, and thus there are four different cases. The quantization rules for these cases are \cite{Braun1993}: 
 \begin{align}
\int_{j_{\textrm{l}}}^{j_{\textrm{r}}}dj p_{j} =  \pi n + 
\begin{cases}
 \frac{\pi}{2}+ \pi (j_{\textrm{r}}-j_{\textrm{l}}), \; j_{\textrm{l}},j_{\textrm{r}} \in U^{-} \\
 -\pi j_{\textrm{l}} ,\quad j_{\textrm{l}} \in U^{-},\; j_{\textrm{r}} \in U^{+} \\
  \pi j_{\textrm{r}} ,\quad j_{\textrm{l}} \in U^{+},\; j_{\textrm{r}} \in U^{-} \\
   \frac{\pi}{2}, \quad j_{\textrm{l}},j_{\textrm{r}} \in U^{+} \\
\end{cases}\,, \label{QuantEq1}
\end{align} 
where $n$ is an integer (potentially negative) and the  WKB momentum function $p_{j}$ is defined as
 \begin{align}
p_{j}\equiv  \arccos\Big(-\frac{ w_{j}}{2 |\mu_{j+1/2}|}\Big)\,.
\end{align} 
The quantization equations (\ref{QuantEq1}) are the transcendental equations which define  quantized values of energy $\varepsilon$, which in turn enters inside the function $p_{j}$ and the turning points $j_{\textrm{l}}, j_{\textrm{r}}$. The equations (\ref{QuantEq1}) are written to $O(\theta^0)$ accuracy, omitting $O(\theta)$ terms. This implies that the function $p_{j}$ is accurate up to $O(\theta)$, since the turning points $j_{\textrm{l}}$ and $j_{\textrm{r}}$ are of order $O(\theta^{-1})$. In general, we expect the energy spacings to be of order $O(\theta)$.

In our case since $j$ runs over the finite interval $[0, n_{m}]$, there are also possible cases when an energy level  $\varepsilon$ 
lacks a left turning point, or a right turning point, or both. Therefore  we have five additional quantization rules: 
 \begin{align}
\int_{j_{\textrm{l}}}^{j_{\textrm{r}}}dj p_{j} =  \pi n - 
\begin{cases}
   p_{0}+\frac{\pi}{4}\,,\;j_{\textrm{l}} =0, \; j_{\textrm{r}} \in U^{+} \\
   p_{0}-\pi j_{\textrm{r}}-\frac{\pi}{4},\; j_{\textrm{l}} =0, \; j_{\textrm{r}} \in U^{-} \\
   p_{n_{\textrm{m}}}+\frac{\pi}{4},\; j_{\textrm{l}} \in U^{+} , \; j_{\textrm{r}}= n_{\textrm{m}} \\
    p_{n_{\textrm{m}}}+\pi j_{\textrm{l}}-\frac{\pi}{4},\;   j_{\textrm{l}} \in U^{-} , \; j_{\textrm{r}}= n_{\textrm{m}} \\
    p_{0}+p_{n_{\textrm{m}}}\,, \; j_{\textrm{l}} =0 , \; j_{\textrm{r}}= n_{\textrm{m}} 
\end{cases}\,.
\end{align} 
We note that, due to the inequalities in (\ref{ChainOfIneq}), the energy $\varepsilon$ does not cross the dispersion $\varepsilon_{1,j}$ in the classically allowed regions, therefore we need not worry about potential singularities in the coefficients $\mu_{j}$ and $w_{j}$. 

Finally, we explain the appearance of flat bands in certain regions of the moir\'{e} bands. First, we note that the moir\'{e} Bloch vector $k$  enters the coefficients $w_{j}$ and $\mu_{j}$ in the combination $k + j b_{\textrm{m}}$.  Now, consider the case in which the left and right turning points $j_{\textrm{l}}$, $j_{\textrm{r}}$ both belong to $U^{-}$ or both belong to $U^{+}$. In either of these two situations, the vector $k$ can be removed from the quantization equations (\ref{QuantEq1}) by the shift $j \to j - k/b_{\textrm{m}}$. Therefore the quantized energies $\varepsilon$ are independent on $k$ up to order $O(\theta^2)$. 
Actually, one can show that, due to the localization of the wave functions in real space, some of these flat bands have an exponentially small bandwidth \cite{ChenVorobevTarn}, similar to the lower-energy bands of the Mathieu equation \cite{PhysRevD.17.498, Dunne:2016qix}.  In the Figure \ref{UpmAndBands1} we can clearly see that flat bands indeed appear in the energy range, where the turning points $j_{\textrm{l}}, j_{\textrm{r}}$ both belong to either  $U^{-}$ or to $U^{+}$ potential functions.

In contrast, if the turning points $j_{\textrm{l}}$ and $j_{\textrm{r}}$ belong to different types of the potential functions, then the quantization equations (\ref{QuantEq1}) depend on the ``unusual" turning point. In that case, the shift  
$j \to j - k/b_{\textrm{m}}$  
removes the  vector $k$ from the left-hand side of (\ref{QuantEq1}), but $k$ reappears explicitly as  
$\pi(n\pm k/b_{\textrm{m}})$  
on the right-hand side of (\ref{QuantEq1}), causing the quantized energy to strongly depend on $k$ and form a zigzag-like energy bands pattern. This behavior is also clearly observed in Figure $\ref{UpmAndBands1}$.

\begin{figure}[h!]
\includegraphics[width=0.48\textwidth]{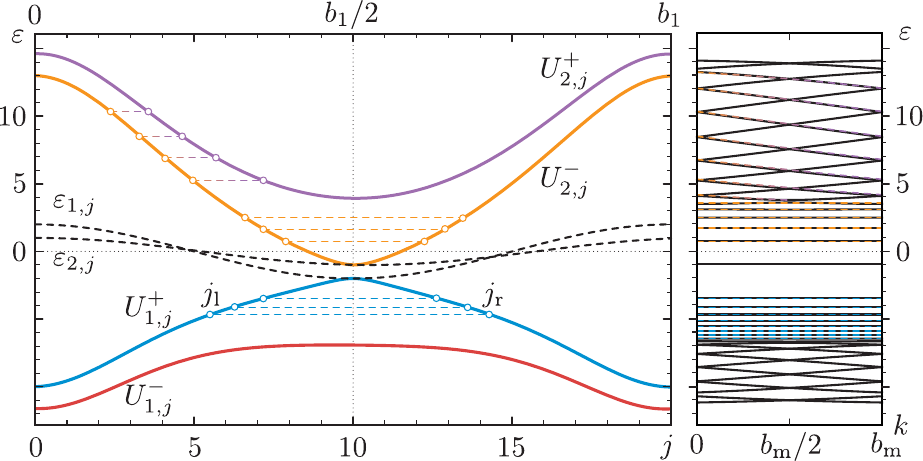}
\caption{(Left) Plot of the potential functions $U^{-}_{1,j}, U^{+}_{1,j}$ and $U^{-}_{2,j}, U^{+}_{2,j}$ (solid colored lines) along with the dispersions $\varepsilon_{1,j}$ and $\varepsilon_{2,j}$ (black dashed lines) for the case $\varepsilon_{1}(p) = 2\cos(p a_{1})$, $\varepsilon_{2}(q) = \cos(q a_{2})$, and $T(p)$ in (\ref{Tpcase1}), with $w = 10$, $\xi/a_{1} = 0.6$, and $\theta = 0.05$. We also plot a few classical energy levels (dashed colored lines) along with their turning points, $j_{\textrm{l}}$ and $j_{\textrm{r}}$. 
(Right) Exact moir\'{e} bands (solid black lines) together with bands obtained from the quantization rules (\ref{QuantEq1}) (dashed colored lines). The energy level near zero is a remnant of the chiral limit flat band and is not captured by (\ref{QuantEq1}).   }
\label{UpmAndBandsWKB}
\end{figure}

In order to demonstrate how the quantization rules (\ref{QuantEq1}) work in practice, we plot Figure \ref{UpmAndBandsWKB} with  parameters: $\varepsilon_{1}(p) = 2\cos(p a_{1})$, $\varepsilon_{2}(q) = \cos(q a_{2})$, and $T(p)$ as in (\ref{Tpcase1}) with $w = 10$, $\xi/a_{1} = 0.6$, and $\theta = 0.05$. On the right side of the figure, we plot the exact moir\'{e} bands (solid black lines) together with the bands obtained from the quantization rules (\ref{QuantEq1}) (dashed colored lines). We see that the accuracy is indeed of order $O(\theta^2)$, and the difference between the exact bands and the WKB ones is almost invisible. We also notice that the energy level near zero is a remnant of the chiral limit flat band, discussed at the end of Section \ref{Sec2} and is not captured by the quantization rules (\ref{QuantEq1}).

\section{Conclusions}
In this article, we showed how the discrete WKB method can reveal the entire structure of the moir\'{e} superchain energy spectrum and explain the emergence of flat bands. By simply constructing the upper and lower potential curves at leading order of accuracy  one can qualitatively understand the structure of the moir\'{e} bands and identify regions where narrow bands arise. One can also use the potential functions to inverse-engineer the desired moir\'{e} bands by tuning the intra- and inter-chain hopping parameters.
We leave a more detailed analysis of the band structure and the stability of the chiral limit flat band, for specific dispersion relations and interchain hopping terms in this model, to future work \cite{ChenVorobevTarn}. We also notice that instead of two chains, one can consider three or more nearby chains of atoms, similar to the 2D multilayers.

It would be interesting to further investigate the role of interactions in the one-dimensional moir\'{e} superchain, depending on the filling factor \cite{PhysRevLett.126.036803}. Finally, we point out that the discrete WKB method is mostly developed for only one-dimensional TTR relations. It would be valuable to generalize these ideas to two-dimensional moir\'{e} superlattices, which could potentially clarify the fundamental principles governing the formation and stability of flat bands in the 2D case.

\vspace{3cm}

\section*{Acknowledgments}  

We are grateful to  Shubhayu Chatterjee, Mitchell Luskin, Tina Kahniashvili, Eslam Khalaf, Igor R. Klebanov and Vlad Kozii for very useful discussions.  We would also like to thank Igor R. Klebanov for his valuable comments on the draft.

\bibliographystyle{ieeetr} 
\bibliography{moire1d}

\begin{thebibliography}{10}

\bibitem{CaoFatemiNature2018}
Y.~Cao, V.~Fatemi, S.~Fang, K.~Watanabe, T.~Taniguchi, E.~Kaxiras, and
  P.~Jarillo-Herrero, ``Unconventional superconductivity in magic-angle
  graphene superlattices,'' {\em Nature}, vol.~556, no.~7699, pp.~43--50, 2018.

\bibitem{CaoFatemiNature2}
Y.~Cao, V.~Fatemi, A.~Demir, S.~Fang, S.~L. Tomarken, J.~Y. Luo, J.~D.
  Sanchez-Yamagishi, K.~Watanabe, T.~Taniguchi, E.~Kaxiras, R.~C. Ashoori, and
  P.~Jarillo-Herrero, ``Correlated insulator behaviour at half-filling in
  magic-angle graphene superlattices,'' {\em Nature}, vol.~556, no.~7699,
  pp.~80--84, 2018.

\bibitem{Yankowitz2018}
M.~Yankowitz, S.~Chen, H.~Polshyn, K.~Watanabe, T.~Taniguchi, D.~Graf, A.~F.
  Young, and C.~R. Dean, ``Tuning superconductivity in twisted bilayer
  graphene,'' {\em arXiv:1808.07865}, 2018.

\bibitem{Li2010}
G.~Li, A.~Luican, J.~M.~B. Lopes~dos Santos, A.~H. Castro~Neto, A.~Reina,
  J.~Kong, and E.~Y. Andrei, ``Observation of van hove singularities in twisted
  graphene layers,'' {\em Nature Physics}, vol.~6, pp.~109--113, February 2010.

\bibitem{bistritzer2011moire}
R.~Bistritzer and A.~H. MacDonald, ``Moir{\'e} bands in twisted double-layer
  graphene,'' {\em Proceedings of the National Academy of Sciences}, vol.~108,
  no.~30, pp.~12233--12237, 2011.

\bibitem{PhysRevB.82.121407}
E.~Su\'arez~Morell, J.~D. Correa, P.~Vargas, M.~Pacheco, and Z.~Barticevic,
  ``Flat bands in slightly twisted bilayer graphene: Tight-binding
  calculations,'' {\em Phys. Rev. B}, vol.~82, p.~121407, Sep 2010.

\bibitem{PhysRevX.8.031089}
H.~C. Po, L.~Zou, A.~Vishwanath, and T.~Senthil, ``Origin of mott insulating
  behavior and superconductivity in twisted bilayer graphene,'' {\em Phys. Rev.
  X}, vol.~8, p.~031089, Sep 2018.

\bibitem{PhysRevB.98.075109}
A.~Thomson, S.~Chatterjee, S.~Sachdev, and M.~S. Scheurer, ``Triangular
  antiferromagnetism on the honeycomb lattice of twisted bilayer graphene,''
  {\em Phys. Rev. B}, vol.~98, p.~075109, Aug 2018.

\bibitem{PhysRevB.98.085435}
L.~Zou, H.~C. Po, A.~Vishwanath, and T.~Senthil, ``Band structure of twisted
  bilayer graphene: Emergent symmetries, commensurate approximants, and wannier
  obstructions,'' {\em Phys. Rev. B}, vol.~98, p.~085435, Aug 2018.

\bibitem{PhysRevB.98.035404}
D.~K. Efimkin and A.~H. MacDonald, ``Helical network model for twisted bilayer
  graphene,'' {\em Phys. Rev. B}, vol.~98, p.~035404, Jul 2018.

\bibitem{PhysRevB.98.045103}
N.~F.~Q. Yuan and L.~Fu, ``Model for the metal-insulator transition in graphene
  superlattices and beyond,'' {\em Phys. Rev. B}, vol.~98, p.~045103, Jul 2018.

\bibitem{PhysRevLett.121.087001}
C.~Xu and L.~Balents, ``Topological superconductivity in twisted multilayer
  graphene,'' {\em Phys. Rev. Lett.}, vol.~121, p.~087001, Aug 2018.

\bibitem{PhysRevB.98.081102}
M.~Ochi, M.~Koshino, and K.~Kuroki, ``Possible correlated insulating states in
  magic-angle twisted bilayer graphene under strongly competing interactions,''
  {\em Phys. Rev. B}, vol.~98, p.~081102, Aug 2018.

\bibitem{PhysRevLett.121.257001}
F.~Wu, A.~H. MacDonald, and I.~Martin, ``Theory of phonon-mediated
  superconductivity in twisted bilayer graphene,'' {\em Phys. Rev. Lett.},
  vol.~121, p.~257001, Dec 2018.

\bibitem{PhysRevB.99.075127}
Y.-H. Zhang, D.~Mao, Y.~Cao, P.~Jarillo-Herrero, and T.~Senthil, ``Nearly flat
  chern bands in moir\'e superlattices,'' {\em Phys. Rev. B}, vol.~99,
  p.~075127, Feb 2019.

\bibitem{PhysRevX.8.031088}
J.~Kang and O.~Vafek, ``Symmetry, maximally localized wannier states, and a
  low-energy model for twisted bilayer graphene narrow bands,'' {\em Phys. Rev.
  X}, vol.~8, p.~031088, Sep 2018.

\bibitem{Pizarro_2019}
J.~M. Pizarro, M.~J. Calderón, and E.~Bascones, ``The nature of correlations
  in the insulating states of twisted bilayer graphene,'' {\em Journal of
  Physics Communications}, vol.~3, p.~035024, mar 2019.

\bibitem{PhysRevX.8.031087}
M.~Koshino, N.~F.~Q. Yuan, T.~Koretsune, M.~Ochi, K.~Kuroki, and L.~Fu,
  ``Maximally localized wannier orbitals and the extended hubbard model for
  twisted bilayer graphene,'' {\em Phys. Rev. X}, vol.~8, p.~031087, Sep 2018.

\bibitem{PhysRevB.98.241407}
D.~M. Kennes, J.~Lischner, and C.~Karrasch, ``Strong correlations and
  $d+\mathit{id}$ superconductivity in twisted bilayer graphene,'' {\em Phys.
  Rev. B}, vol.~98, p.~241407, Dec 2018.

\bibitem{PhysRevX.8.041041}
H.~Isobe, N.~F.~Q. Yuan, and L.~Fu, ``Unconventional superconductivity and
  density waves in twisted bilayer graphene,'' {\em Phys. Rev. X}, vol.~8,
  p.~041041, Dec 2018.

\bibitem{PhysRevB.98.235158}
L.~Rademaker and P.~Mellado, ``Charge-transfer insulation in twisted bilayer
  graphene,'' {\em Phys. Rev. B}, vol.~98, p.~235158, Dec 2018.

\bibitem{PhysRevB.98.220504}
T.~J. Peltonen, R.~Ojaj\"arvi, and T.~T. Heikkil\"a, ``Mean-field theory for
  superconductivity in twisted bilayer graphene,'' {\em Phys. Rev. B}, vol.~98,
  p.~220504, Dec 2018.

\bibitem{PhysRevB.99.144507}
V.~Kozii, H.~Isobe, J.~W.~F. Venderbos, and L.~Fu, ``Nematic superconductivity
  stabilized by density wave fluctuations: Possible application to twisted
  bilayer graphene,'' {\em Phys. Rev. B}, vol.~99, p.~144507, Apr 2019.

\bibitem{PhysRevB.106.235157}
V.~Kozii, M.~P. Zaletel, and N.~Bultinck, ``Spin-triplet superconductivity from
  intervalley goldstone modes in magic-angle graphene,'' {\em Phys. Rev. B},
  vol.~106, p.~235157, Dec 2022.

\bibitem{PhysRevLett.123.036401}
Z.~Song, Z.~Wang, W.~Shi, G.~Li, C.~Fang, and B.~A. Bernevig, ``All magic
  angles in twisted bilayer graphene are topological,'' {\em Phys. Rev. Lett.},
  vol.~123, p.~036401, Jul 2019.

\bibitem{PhysRevB.99.035111}
K.~Hejazi, C.~Liu, H.~Shapourian, X.~Chen, and L.~Balents, ``Multiple
  topological transitions in twisted bilayer graphene near the first magic
  angle,'' {\em Phys. Rev. B}, vol.~99, p.~035111, Jan 2019.

\bibitem{PhysRevB.99.195455}
H.~C. Po, L.~Zou, T.~Senthil, and A.~Vishwanath, ``Faithful tight-binding
  models and fragile topology of magic-angle bilayer graphene,'' {\em Phys.
  Rev. B}, vol.~99, p.~195455, May 2019.

\bibitem{PhysRevLett.108.216802}
P.~San-Jose, J.~Gonz\'alez, and F.~Guinea, ``Non-abelian gauge potentials in
  graphene bilayers,'' {\em Phys. Rev. Lett.}, vol.~108, p.~216802, May 2012.

\bibitem{PhysRevLett.122.106405}
G.~Tarnopolsky, A.~J. Kruchkov, and A.~Vishwanath, ``Origin of magic angles in
  twisted bilayer graphene,'' {\em Phys. Rev. Lett.}, vol.~122, p.~106405, Mar
  2019.

\bibitem{PhysRevResearch.2.023237}
P.~J. Ledwith, G.~Tarnopolsky, E.~Khalaf, and A.~Vishwanath, ``Fractional chern
  insulator states in twisted bilayer graphene: An analytical approach,'' {\em
  Phys. Rev. Res.}, vol.~2, p.~023237, May 2020.

\bibitem{PhysRevB.103.155150}
F.~K. Popov and A.~Milekhin, ``Hidden wave function of twisted bilayer
  graphene: The flat band as a landau level,'' {\em Phys. Rev. B}, vol.~103,
  p.~155150, Apr 2021.

\bibitem{PhysRevResearch.3.023155}
J.~Wang, Y.~Zheng, A.~J. Millis, and J.~Cano, ``Chiral approximation to twisted
  bilayer graphene: Exact intravalley inversion symmetry, nodal structure, and
  implications for higher magic angles,'' {\em Phys. Rev. Res.}, vol.~3,
  p.~023155, May 2021.

\bibitem{PhysRevLett.127.246403}
J.~Wang, J.~Cano, A.~J. Millis, Z.~Liu, and B.~Yang, ``Exact landau level
  description of geometry and interaction in a flatband,'' {\em Phys. Rev.
  Lett.}, vol.~127, p.~246403, Dec 2021.

\bibitem{PhysRevX.13.021012}
Y.~Sheffer, R.~Queiroz, and A.~Stern, ``Symmetries as the guiding principle for
  flattening bands of dirac fermions,'' {\em Phys. Rev. X}, vol.~13, p.~021012,
  Apr 2023.

\bibitem{PhysRevB.100.085109}
E.~Khalaf, A.~J. Kruchkov, G.~Tarnopolsky, and A.~Vishwanath, ``Magic angle
  hierarchy in twisted graphene multilayers,'' {\em Phys. Rev. B}, vol.~100,
  p.~085109, Aug 2019.

\bibitem{PhysRevLett.123.026402}
C.~Mora, N.~Regnault, and B.~A. Bernevig, ``Flatbands and perfect metal in
  trilayer moir\'e graphene,'' {\em Phys. Rev. Lett.}, vol.~123, p.~026402, Jul
  2019.

\bibitem{CeaWaletGuinea2019}
T.~Cea, N.~R. Walet, and F.~Guinea, ``Twists and the electronic structure of
  graphitic materials,'' {\em Nano Letters}, vol.~19, pp.~8683--8689, 12 2019.

\bibitem{PhysRevLett.125.116404}
Z.~Zhu, S.~Carr, D.~Massatt, M.~Luskin, and E.~Kaxiras, ``Twisted trilayer
  graphene: A precisely tunable platform for correlated electrons,'' {\em Phys.
  Rev. Lett.}, vol.~125, p.~116404, Sep 2020.

\bibitem{lin2022energetic}
X.~Lin, C.~Li, K.~Su, and J.~Ni, ``Energetic stability and spatial
  inhomogeneity in the local electronic structure of relaxed twisted trilayer
  graphene,'' {\em Physical Review B}, vol.~106, no.~7, p.~075423, 2022.

\bibitem{ma2023doubled}
Z.~Ma, S.~Li, M.~Lu, D.-H. Xu, J.-H. Gao, and X.~Xie, ``Doubled moir{\'e} flat
  bands in double-twisted few-layer graphite,'' {\em Science China Physics,
  Mechanics \& Astronomy}, vol.~66, no.~2, p.~227211, 2023.

\bibitem{PhysRevB.105.195422}
M.~Liang, M.-M. Xiao, Z.~Ma, and J.-H. Gao, ``Moir\'e band structures of the
  double twisted few-layer graphene,'' {\em Phys. Rev. B}, vol.~105, p.~195422,
  May 2022.

\bibitem{PhysRevLett.128.176404}
P.~J. Ledwith, A.~Vishwanath, and E.~Khalaf, ``Family of ideal chern flatbands
  with arbitrary chern number in chiral twisted graphene multilayers,'' {\em
  Phys. Rev. Lett.}, vol.~128, p.~176404, Apr 2022.

\bibitem{PhysRevB.107.125423}
Y.~Mao, D.~Guerci, and C.~Mora, ``Supermoir\'e low-energy effective theory of
  twisted trilayer graphene,'' {\em Phys. Rev. B}, vol.~107, p.~125423, Mar
  2023.

\bibitem{PhysRevB.108.L081124}
F.~K. Popov and G.~Tarnopolsky, ``Magic angles in equal-twist trilayer
  graphene,'' {\em Phys. Rev. B}, vol.~108, p.~L081124, Aug 2023.

\bibitem{PhysRevResearch.6.L022025}
D.~Guerci, Y.~Mao, and C.~Mora, ``Chern mosaic and ideal flat bands in
  equal-twist trilayer graphene,'' {\em Phys. Rev. Res.}, vol.~6, p.~L022025,
  Apr 2024.

\bibitem{Devakul2023}
T.~Devakul, P.~J. Ledwith, L.-Q. Xia, A.~Uri, S.~C. de~la Barrera,
  P.~Jarillo-Herrero, and L.~Fu, ``Magic-angle helical trilayer graphene,''
  {\em Science Advances}, vol.~9, no.~36, p.~eadi6063, 2023.

\bibitem{PhysRevX.13.041007}
N.~Nakatsuji, T.~Kawakami, and M.~Koshino, ``Multiscale lattice relaxation in
  general twisted trilayer graphenes,'' {\em Phys. Rev. X}, vol.~13, p.~041007,
  Oct 2023.

\bibitem{PhysRevResearch.5.043079}
F.~K. Popov and G.~Tarnopolsky, ``Magic angle butterfly in twisted trilayer
  graphene,'' {\em Phys. Rev. Res.}, vol.~5, p.~043079, Oct 2023.

\bibitem{PhysRevB.109.205411}
D.~Guerci, Y.~Mao, and C.~Mora, ``Nature of even and odd magic angles in
  helical twisted trilayer graphene,'' {\em Phys. Rev. B}, vol.~109, p.~205411,
  May 2024.

\bibitem{Canc_s_2017}
E.~Cancès, P.~Cazeaux, and M.~Luskin, ``Generalized kubo formulas for the
  transport properties of incommensurate 2d atomic heterostructures,'' {\em
  Journal of Mathematical Physics}, vol.~58, June 2017.

\bibitem{PhysRevResearch.2.033162}
S.~Carr, D.~Massatt, M.~Luskin, and E.~Kaxiras, ``Duality between atomic
  configurations and bloch states in twistronic materials,'' {\em Phys. Rev.
  Res.}, vol.~2, p.~033162, Jul 2020.

\bibitem{PhysRevB.101.041112}
M.~Fujimoto, H.~Koschke, and M.~Koshino, ``Topological charge pumping by a
  sliding moir\'e pattern,'' {\em Phys. Rev. B}, vol.~101, p.~041112, Jan 2020.

\bibitem{PhysRevB.101.041113}
Y.~Su and S.-Z. Lin, ``Topological sliding moir\'e heterostructure,'' {\em
  Phys. Rev. B}, vol.~101, p.~041113, Jan 2020.

\bibitem{Tritsaris2021}
G.~A. Tritsaris, S.~Carr, and G.~R. Schleder, ``Computational design of moiré
  assemblies aided by artificial intelligence,'' {\em Applied Physics Reviews},
  vol.~8, p.~031401, 07 2021.

\bibitem{Gon_alves_2022}
M.~Gonçalves, B.~Amorim, E.~Castro, and P.~Ribeiro, ``Hidden dualities in 1d
  quasiperiodic lattice models,'' {\em SciPost Physics}, vol.~13, Sept. 2022.

\bibitem{PhysRevResearch.4.043224}
D.~Liu, M.~Luskin, and S.~Carr, ``Seeing moir\'e: Convolutional network
  learning applied to twistronics,'' {\em Phys. Rev. Res.}, vol.~4, p.~043224,
  Dec 2022.

\bibitem{PhysRevB.106.075420}
Q.~Gao and E.~Khalaf, ``Symmetry origin of lattice vibration modes in twisted
  multilayer graphene: Phasons versus moir\'e phonons,'' {\em Phys. Rev. B},
  vol.~106, p.~075420, Aug 2022.

\bibitem{PhysRevResearch.4.L032031}
D.~X. Nguyen, X.~Letartre, E.~Drouard, P.~Viktorovitch, H.~C. Nguyen, and H.~S.
  Nguyen, ``Magic configurations in moir\'e superlattice of bilayer photonic
  crystals: Almost-perfect flatbands and unconventional localization,'' {\em
  Phys. Rev. Res.}, vol.~4, p.~L032031, Aug 2022.

\bibitem{Dams2023}
D.~Dams, D.~Beutel, X.~Garcia-Santiago, C.~Rockstuhl, and R.~Alaee, ``Moir\'{e}
  flat bands in strongly coupled atomic arrays,'' {\em Opt. Mater. Express},
  vol.~13, pp.~2003--2019, Jul 2023.

\bibitem{Schleder_2023}
G.~R. Schleder, M.~Pizzochero, and E.~Kaxiras, ``One-dimensional moiré physics
  and chemistry in heterostrained bilayer graphene,'' {\em The Journal of
  Physical Chemistry Letters}, vol.~14, p.~8853–8858, Sept. 2023.

\bibitem{PhysRevB.107.224206}
D.~Vu and S.~Das~Sarma, ``Generic mobility edges in several classes of
  duality-breaking one-dimensional quasiperiodic potentials,'' {\em Phys. Rev.
  B}, vol.~107, p.~224206, Jun 2023.

\bibitem{Braun1993}
P.~A. Braun, ``Discrete semiclassical methods in the theory of rydberg atoms in
  external fields,'' {\em Rev. Mod. Phys.}, vol.~65, pp.~115--161, Jan 1993.

\bibitem{PhysRevLett.125.166803}
A.~Timmel and E.~J. Mele, ``Dirac-harper theory for one-dimensional moir\'e
  superlattices,'' {\em Phys. Rev. Lett.}, vol.~125, p.~166803, Oct 2020.

\bibitem{Becker2024}
S.~Becker and J.~Wittsten, ``Semiclassical quantization conditions in strained
  moiré lattices,'' {\em Communications in Mathematical Physics}, vol.~405,
  no.~9, p.~218, 2024.

\bibitem{PhysRevB.107.235143}
E.~Andrade, F.~L\'opez-Ur\'{\i}as, and G.~G. Naumis, ``Topological origin of
  flat bands as pseudo-landau levels in uniaxial strained graphene nanoribbons
  and induced magnetic ordering due to electron-electron interactions,'' {\em
  Phys. Rev. B}, vol.~107, p.~235143, Jun 2023.

\bibitem{Gantmacher1961OscillationMA}
F.~R. Gantmacher and M.~Krein, ``Oscillation matrices and kernels and small
  vibrations of mechanical systems,'' 1961.

\bibitem{Harper_1955}
P.~G. Harper, ``Single band motion of conduction electrons in a uniform
  magnetic field,'' {\em Proceedings of the Physical Society. Section A},
  vol.~68, p.~874, oct 1955.

\bibitem{SchultenGordon1975}
K.~Schulten and R.~G. Gordon, ``Exact recursive evaluation of 3j‐ and
  6j‐coefficients for quantum‐mechanical coupling of angular momenta,''
  {\em Journal of Mathematical Physics}, vol.~16, pp.~1961--1970, 10 1975.

\bibitem{Sazonov1978}
V.~N. Sazonov, ``Quasiclassical theory of the excitation of a quantum nonlinear
  oscillator,'' {\em Theoretical and Mathematical Physics}, vol.~35, no.~3,
  pp.~514--520, 1978.

\bibitem{Braun1978}
P.~A. Braun, ``Wkb method for three-term recursion relations and quasienergies
  of an anharmonic oscillator,'' {\em Theoretical and Mathematical Physics},
  vol.~37, no.~3, pp.~1070--1081, 1978.

\bibitem{Braun1979}
P.~A. Braun, ``Quasienergies of an anharmonic oscillator in parametric
  resonance,'' {\em Theoretical and Mathematical Physics}, vol.~41, no.~3,
  pp.~1060--1066, 1979.

\bibitem{PhysRevLett.43.1954}
M.~Y. Azbel, ``Quantum particle in one-dimensional potentials with
  incommensurate periods,'' {\em Phys. Rev. Lett.}, vol.~43, pp.~1954--1957,
  Dec 1979.

\bibitem{Wilkinson1984}
M.~Wilkinson, ``Critical properties of electron eigenstates in incommensurate
  systems,'' {\em Proceedings of the Royal Society of London. Series A,
  Mathematical and Physical Sciences}, vol.~391, no.~1801, pp.~305--350, 1984.

\bibitem{ChenVorobevTarn}
Y.~Chen, D.~Vorobev, and G.~M. Tarnopolsky {\em In preparation}.

\bibitem{PhysRevD.17.498}
H.~Neuberger, ``Semiclassical calculation of the energy dispersion relation in
  the valence band of the quantum pendulum,'' {\em Phys. Rev. D}, vol.~17,
  pp.~498--506, Jan 1978.

\bibitem{Dunne:2016qix}
G.~V. Dunne and M.~Unsal, ``{WKB and Resurgence in the Mathieu Equation},'' 3
  2016.

\bibitem{PhysRevLett.126.036803}
D.~Vu and S.~Das~Sarma, ``Moir\'e versus mott: Incommensuration and interaction
  in one-dimensional bichromatic lattices,'' {\em Phys. Rev. Lett.}, vol.~126,
  p.~036803, Jan 2021.

\end{thebibliography}

\end{document}